\documentstyle[11pt]{article}
\renewcommand{\cite}[1]{\ref{#1}}

\newcommand{\beq}{\begin{equation}}
\newcommand{\eeq}{\end{equation}}
\newcommand{\beqa}{\begin{eqnarray}}
\newcommand{\eeqa}{\end{eqnarray}}
\newcommand{\bcent}{\begin{center}}
\newcommand{\ecent}{\end{center}}

\newcommand{\reflef}{(\ref}
\newcommand{\Hbar}{\bar{\rm H}}

\begin{document}

\baselineskip=0.4cm
\begin{center}
{\Large\bf What can we do on gravity with antihydrogens?\footnote{Delivered at The International Workshop on JHF 
Science (JHF98), KEK, Tsukuba, Japan, March 4-7, 1998.}}\vspace{.6em}\\
Yasunori Fujii\footnote{E-mail address: fujii@handy.n-fukushi.ac.jp}\vspace{.6em}\\
Nihon Fukushi University, Handa, 475-0012\ Japan\\
and\\
ICRR, University of Tokyo, Tanashi, Tokyo, 188-8502\ Japan
\end{center}

\bigskip
\baselineskip=0.5cm
\bcent
{\large\bf Abstract}\\
\bigskip
\begin{minipage}{13cm}
We discuss the implications of the proposed gravitational redshift
experiment on antihydrogens.  We show that the result should be the
same as on hydrogens in spite of different free-fall 
accelerations (WEP violation) which may occur if there is a vector
fifth-force field.  We emphasize the experiment is unique in the sense
that it tests the Equivalence Principle expressed in its ultimate 
form, proposed to be called UEP, directly without being disturbed by
the effect of  possible presence of a scalar fifth-force field.
\end{minipage}
\ecent
\bigskip

Trying to see how antiparticles respond to gravity has a long history, but the efforts have been hampered mainly by the presence of the electrostatic force which overwhelms the gravitational force.  In this respect antihydrogens, if sufficient number of them are available [\cite{exp}], should have an obvious advantage.  If gravity is precisely what Einstein envisioned, antiparticles should fall with the same acceleration as particles, a manifestation of the Equivalence Principle (EP).  Expecting violation of this sacred ingredient in General Relativity (GR) is, however, not entirely a trivial idea because almost any of the  modern versions of unification program suggest possible occurrence of  hitherto unknown fields which might mimic the ordinary gravitational field.

In fact there are some fields that couple to the ``ordinary" matter fields, like quarks, leptons and gauge fields in the standard model, basically as weakly as the gravitational coupling.  But they have nothing to do with the spacetime geometry at least in 4 dimensions.  For this reason they might be called ``non-geometric gravitational fields" (NGGFs).  They are simply part of ``matter fields," which were sharply distinguished by Einstein from the metric field.   Some of them might be as heavy as the Planck mass hence remaining permanently unobservable, but  some others can be extremely light or nearly massless, thus manifesting themselves as  ``fifth forces" [\cite{fifth}].

It should be pointed out that fields of this kind fail generically to  observe composition-independence in their matter coupling, hence violating universal free-fall (UFF).  This UFF is one of the aspects of EP.  To emphasize a special position of UFF in the more general concept of EP, people give this  a special name; Weak Equivalence Principle (WEP).  Put in the other way around, WEP is a privilege enjoyed only by a force endowed with spacetime geometry.  No other forces, including those due to NGGFs, share this luxury.  For this reason, testing WEP is a way to probe a fifth force.

Even combined with another way of probing departure from a purely inverse-square law, no solid evidence for the fifth force has been reported.  This does not imply, however, that its possible presence has been ruled out.  Searching it below the available upper bounds is still going on in various ways.  In this context we try to see how $\Hbar$ will respond to a fifth force.

Among many possibilities on the type of a fifth force, we are particularly interested in a vector field, because this is the only choice which makes a difference between H and $\Hbar$.  On  Earth's surface, a vector force will repel H but attract $\Hbar$, contributing the accelerations $-g_v$ and $g_v$, respectively, though the exact  size depends on the strength of the coupling and the force-range.  As a consequence we have $g_{\rm H}=g_t-g_v$ and $g_{\bar{\rm H}}=g_t +g_v$, hence resulting in WEP violation, where $g_t$ is the ordinary acceleration coming from the tensor gravitational force.

We now turn to another type of proposed experiment, gravitational redshift [\cite{redsh}].  What do we expect to get by measuring gravitational redshift of $\Hbar$ and comparing it with that of H?

Suppose H or $\Hbar$ emits radiation received by another object, which may or may not be another H atom located at another place higher by $h$.  We may consider $2S$-$1S$ transition  or  radiation from the hyperfine splitting measured by a cavity.

The text-book derivation of the gravitational redshift is a combination of the two steps: (A) Based on the static nature of the setting, one derives 
\beq
\Delta t_{\rm em}=\Delta t_{\rm rec}, 
\label{kk2_4}
\eeq
on the intervals of the {\em coordinate} times for one cycle of the oscillation for emitter and receiver.  (B) At each side one also derives \beq
\Delta\tau =\sqrt{-g_{00}}\Delta t, 
\label{kk2_5}
\eeq
to correlate the {\em proper} time difference $\Delta\tau$ and the local {\em coordinate} time difference $\Delta t$.  We then obtain
\beq
Z\equiv\frac{\Delta\omega}{\omega}=-c^{-2}\Delta U= -g c^{-2}h,
\label{kk2_6}
\eeq
where $U(z)= gz$ is the Newtonian potential divided by mass, as dictated from $-g_{00}=1+2c^{-2}U(z)$.

Of crucial importance is to recognize that  laws like the ``frequency
condition'' $\Delta E=\hbar\omega$ in quantum mechanics can be applied
only in the {\em freely falling, local inertial frames,} with the
proper time $\tau$ a time variable. Exact details of quantum mechanics in curved spacetime might be complicated, but all we need for the present purpose is \reflef{kk2_5}) that embodies what we are going to call Ultimate Einstein Equivalence Principle (UEP) in the simplest form.  We have derived \reflef{kk2_6}) independently of whether the radiation is emitted by H or $\Hbar$.  It then follows that {\em no difference} is expected between the redshifts of H and $\Hbar$:
\beq
Z_{\rm H} =Z_{\Hbar}.
\label{kk2_7}
\eeq
We add that this conclusion holds true even for the so-called null-redshift experiments.

What is UEP?    People now talk about
many kinds of, or many aspects of EP.  As was mentioned, WEP is the
same as UFF,  a ``phenomenological" law.  To understand this  ``mysterious" law, Einstein came up with  a geometrical origin behind it.   Clifford Will proposed to call this theoretical law at a more fundamental level Einstain Equivalence Principle (EEP), which says   that {\em any} gravitational phenomena are consequences of curved spacetime [\cite{will}].

Using the word ``any"  was important when he tried
to sort out  some of the  alternative theories, mainly in the
astronomical distances.  Since the fifth force 
idea was proposed, however, it is almost agreed that, at least in the
medium-range distances, we should include NGGFs that participate
gravitational phenomena in a wider sense; gravitational phenomena may
not be consequences  of pure spacetime geometry.  Will's EEP as it stands seems too limited in scope as a basis of discussing  unification-oriented physics of gravity.

On the other hand, we have no intention to degrade Einstein.  On the contrary, even extending the content of gravitational phenomena, we
still want to maintain the heart of GR.  That is UEP.  Einstein elevated his idea of ``Einstein's lift" to a more fundamental
law that gravity can always be eliminated by a coordinate
transformation.  In the
final theory of GR, EP was given a more abstract mathematical
expression that {\em tangential spacetime at a world point on
Riemanian manifold is Mikowskian}, representing a local inertial
frame.  To make a distinction from Will's EEP, we propose to call this UEP.  The same thing was called simply EP in Weinberg's book [\cite{wb}].  He used
the term Strong Equivalence Principle (SEP) referring his EP stated
above applied to all physical laws beyond simple mechanical systems,
though Will prefers to choose the same name SEP implying UFF applied to those objects with non-negligible amount of gravitational binding energies, namely extending WEP to higher orders in $G$.

We reiterate that UEP played a crucial role in deriving \reflef{kk2_6}).  To discuss the frequency we might have used the coordinate times at both sides, for example, then arriving at $\Delta \omega =0$ due to \reflef{kk2_4}).  Other reference frames could have also been used, as will be discussed later.  In this sense, redshift experiments can be interpreted as testing UEP.

A number of alternative derivations of the same result have been proposed in the literature; based only on EP and Special Relativity, or relying only on the energy conservation law.  It may sound as if UEP were not relevant.  This is, however, only superficial.  The former argument defines  the physical frequency explicitly in the freely falling systems, which are compared with each other by means of a Lorentz transformation.

In the latter derivation, on the other hand, the frequency condition $\pm \omega = E^* -E$ is used as a conservation law, corrected for the gravitational interaction.  Late John Bell emphasized that this is done without ``philosophying" about spacetime [\cite{bell}].  But he essentially  rephrased the above argument based on UEP in the first-order perturbation theory.  These derivations might be favored by those who are not familiar with or abhor the concept of spacetime geometry.  But suppose breakdown of \reflef{kk2_7}) is in fact verified experimentally.  Are we relieved by thinking that only some theoretical approach of limited applicability is rejected?  Wouldn't it  better to recognize how serious it would be if the heart of GR is at stake?

To illustrate how easy it is to arrive at a wrong result if we choose
a wrong coordinate frame, let us give  an example. We go to the freely
falling coordinate systems which are {\em different} for H and $\Hbar$ if $g_{v}\neq 0$.  By assuming that these are the correct frames in which the frequency condition is used, one substitutes $g_{\rm H}$ and $g_{\bar{\rm H}}$ into \reflef{kk2_6}), thus arriving at
\beq
Z_{\rm H}=-g_{\rm H} c^{-2}\Delta z,\quad\mbox{and}\quad Z_{\Hbar}=-g_{\bar{\rm H}} c^{-2}\Delta z,
\label{ap3}
\eeq
resulting in different redshifts for H and $\Hbar$.

The argument is unjustified, however, because the freely falling systems are {\em not} the local inertial frames; $g_{t}$ remains nonzero to balance $\pm g_{v}$.  Spacetime is still curved, not allowing to apply UEP.  The correct local inertial frame is {\em common} to both of H and $\Hbar$, hence no difference between $Z_{\rm H}$ and $Z_{\Hbar}$; $g$ in \reflef{kk2_6}) should be replaced by $g_{t}$.

Let us also give  another example showing how serious it would be if a difference comes out.  Try to find a {\em geometrical} derivation of 
\beq
Z_{\rm H}= \alpha Z_{\Hbar},
\label{ap4}
\eeq
with $\alpha\neq 1$. This could be achieved if $-g_{00}$ in \reflef{kk2_5}) for $\Hbar$ is replaced by $(-g_{00})^{\alpha}$: 
\beq
-\overline{g}_{00}=(-g_{00})^{\alpha}.
\label{ap5}
\eeq
But we are now preparing a {\em different} spacetime for
$\Hbar$.  At the sacrifice is the simple picture that everything lives in a single and common spacetime.  Spacetime approach loses much of its predicting power.

So far we have not taken into account the fact that even in the local
inertial frame we still have the {\em residual} vector force; repulsive for
H, while attractive for $\Hbar$.  However, the vector  field is a
gauge field that couples to some conserved charge, baryon number or
lepton number, for example.  In any case the charge is the same for
the ground state and the excited states.  We have no effect for the {\em difference} between the ground state and
the excited state.  For this reason the transition frequency, and
hence the result \reflef{kk2_7})  remains unaffected.   This also implies that the ground state and an excited state should fall with different accelerations.  Violation of WEP in this sense is not detected by the redshift experiment.

Summarizing,  no difference should be expected between
redshifts of H and $\Hbar$, in spite of the fact that they might fall
with different accelerations, implying WEP violation.  If any
difference is detected in the redshifts of H and $\Hbar$, we should
suspect some violation at a higher level, breakdown of UEP, or
something even more serious.

How important is UEP?  From a theoretical point of view, it is of crucial importance and convenience.  It helps us to find physical laws in the presence of gravity out from those in its absence.  Notably one can apply such powerful rules as $\eta_{\mu\nu}\rightarrow g_{\mu\nu}, \partial_{\mu}\rightarrow \nabla_{\mu}$, sometimes called the ``comma-goes-to-semicolon rule" [\cite{telb}].  Electromagnetism and quantum mechanics, for example, in the gravitational field are established in this way. No modern attempts to build a theoretical model of unification has been made without the use of UEP.

From an experimental point of view, we emphasize that this redshift experiment comparing H and $\Hbar$ is quite unique in that we can test UEP directly without being disturbed by the possible presence of a scalar fifth-force field.  Obviously it is unaffected by a scalar field, because the scalar field, if any, should couple equally to H and $\Hbar$.  What we point out here is that almost any other experiments are not immune from a scalar field in principle, putting aside a quantitative discussion, for the moment.  One might hear the argument accepted rather widely that the behavior of the light in the gravitational field, the light deflection or Shapiro's time delay, for example, is left unaffected by a scalar field.  This is, however, based on the assumption that the scalar field couples to the {\em trace} of the energy-momentum tensor, as in Brans-Dicke model, a feature shared by few of the theoretical models of unification.

The author thanks Takashi Ishikawa, Kenji Fukushima, Ephraim Fishcbach and John Eades for valuable discussions.
\bigskip
\newpage
\bcent
{\large\bf References}
\ecent
\bigskip
\baselineskip=0.4cm

\begin{enumerate}
\item\label{exp}See, for example, M.Charlton, J.Eades, D.Horvath, R.J.Hughes and C.Zimmermann, Phys. Rep. {\bf t241}(1994)65, and papers cited therein.
\item\label{fifth}See, for recent development, Y. Fujii, Prog. Theor. Phys. {\bf 99}(1998) 599, (gr-qc/9708010).
\item\label{redsh}R. Hughes, M. Holzscheiter, J. Modern Optics, {\bf 39}(1992)263;  R.J. Hughes, Hyperfine Int.{\bf 76}(1993)3, and papers cited therein.  See also M.M. Nieto and T. Goldman,  Phys. Rep.{\bf 205}(1991)221.
\item\label{will}C. Will, Int. J. Mod. Phys.{\bf D1}(1992) 13: {\sl Theory and Experiments in Gravitational Physics}, rev. ed., Cambridge University Press, (1993).
\item\label{wb}S. Weinberg, {\sl Gravitation and Cosmology}, John Wiley, (1972).
\item\label{bell}J.S. Bell, in {\sl Fundamental Symmetries}, ed. P. Bloch, P. Pavlopoulos and R. Klapisch, 1987, Plenum Press.
\item\label{telb}C.W. Misner, K.S. Thorn and
J.A. Wheeler, {\sl Gravitation}, p. 387, W.H. Freeman, (1973).
\end{enumerate}

\end{document}